\documentclass[useAMS,usenatbib]{mn2e}
\usepackage[dvipdfm]{graphicx,color}

\title[Condensation of a radiative gas layer]
{Self-similar solutions for the dynamical condensation of a 
radiative gas layer}
\author[Kazunari Iwasaki and Toru Tsuribe]
       {Kazunari Iwasaki$^{1}$\thanks{E-mail: iwasaki@vega.ess.sci.osaka-u.ac.jp}
      and Toru  Tsuribe$^{1}$ \\
$^{1}$ Department of Earth and Space Science, Osaka University, Machikaneyama 1-1 
   Toyonaka, Osaka 560-0043, Japan}

\begin{document}

\date{}
\maketitle


\begin{abstract}
A new self-similar solution describing the dynamical condensation of 
a radiative gas is investigated under a plane-parallel geometry.
The dynamical condensation is caused by thermal instability. 
The solution is applicable to generic flow with
a net cooling rate per unit volume and time 
$\propto \rho^2 T^\alpha$, 
where $\rho$, $T$ and $\alpha$ are
density, temperature and a free parameter, respectively.
Given $\alpha$, a family of self-similar solutions with one parameter $\eta$ is 
found in which the central density and pressure 
evolve as follows: $\rho(x=0,t)\propto (t_\mathrm{c}-t)^{-\eta/(2-\alpha)}$ and 
$P(x=0,t)\propto (t_\mathrm{c}-t)^{(1-\eta)/(1-\alpha)}$, where $t_\mathrm{c}$ is 
an epoch when the central density becomes infinite.
For $\eta\sim 0$, the solution describes the isochoric mode,
whereas for $\eta\sim1$, the solution describes the isobaric mode. 
The self-similar solutions exist in the range between 
the two limits; that is, for $0<\eta<1$. 
No self-similar solution is found for $\alpha>1$.
We compare the obtained self-similar solutions with the 
results of one-dimensional hydrodynamical simulations.
In a converging flow, the results of 
the numerical simulations
agree well with the self-similar solutions in 
the high-density limit.
Our self-similar solutions are applicable to the 
formation of interstellar clouds (HI cloud and 
molecular cloud) by thermal instability.
\end{abstract}

\section{Introduction}
Thermal instability (TI) is an important physical process in 
astrophysical environments that are subject to radiative cooling.
\citet{FR85} investigated the effect of TI caused by Bremsstrahlung 
on structure formation at galactic and subgalactic scales.
TI also plays an important role in the interstellar medium (ISM). 
The ISM consists of a diffuse high-temperature phase (warm 
neutral medium, or WNM) and a clumpy low-temperature phase 
(cold neutral medium, 
or CNM), in supersonic turbulence.
It is known that the ISM is thermally unstable
in the temperature range between 
these the two stable phases; that is, in the range 
$T\sim 300-6000$ K \citep{DM72}.
Koyama \& Inutsuka (2000, 2002) suggested that 
the structure of the ISM is provided by the phase transition 
that is caused by TI that occurs in a shock-compressed region.

The basic properties of TI were investigated by \citet{F65},
who investigated the TI of a spatially uniform gas
using linear analysis for the unperturbed state in thermal equilibrium.
The net cooling function per unit volume and time was assumed 
to be $\Lambda_0\rho^2T^\alpha - \Gamma_0 \rho$.
\citet{F65} showed that gas is isobarically unstable for $\alpha<1$ and 
is isochorically unstable for $\alpha<0$.
In thermal non-equilibrium, however, where
the heating and the cooling rates are not equal in the unperturbed state,
the stability condition of TI is different from that 
in thermal equilibrium. 
This situation is, for example, generated in a shock-compressed region.
\citet{B86} derived a general criterion for the TI
of gas in a non-equilibrium state.
A flow dominated by cooling is isobarically unstable for
$\alpha\la2$ and is also isochorically unstable for $\alpha\la 1$.
Therefore, the phase with $T\sim 300-6000$K is both isobarically 
and isochorically unstable. 

The non-linear evolution of TI has been investigated by many authors 
\citep{KI02,AH05,HA07,H06,V07} using multi-dimensional simulations.
However, the nonlinear behaviour has only been 
investigated analytically 
by a few authors, using some simplifications 
in an attempt to gain a deeper insight into the nature of TI. 
\citet{M89} 
investigated analytic solutions under the isobaric approximation.
However, this approximation is valid only in the limit of 
short or intermediate-scale.

In this paper, a new semi-analytic model for the non-linear hydrodynamical 
evolution of TI for a thermally 
non-equilibrium gas is considered without the isobaric approximation. 
Once TI occurs, the density increases drastically 
until a thermally stable phase it reached. 
During this dynamical condensation, the gas is expected 
to lose its memory of the initial condition, and 
is expected to behave in a self-similar manner 
\citep{ZR67,LL59}.
In particularly, in star formation, 
self-similar (S-S) solutions
for a runaway collapse of a self-gravitating 
gas have been investigate
by many authors \citep{P69,L69,S77,H77,WS85,BL95}. 
In this paper, S-S solutions describing dynamical condensation by TI 
are investigated by assuming a simple cooling rate without self-gravity.
A family of S-S solutions which have two asymptotic limits 
(the isobaric and the isochoric modes) is presented.

In Section \ref{formulation}, we derive the S-S equations, and describe 
the mathematical characteristic 
of our S-S equations as well as the numerical methods.
In Section \ref{result}, our S-S solutions are presented and their properties 
are discussed.
In Section \ref{discuss}, the S-S solutions are 
compared with the results of one-dimensional 
numerical simulations.
The astrophysical implications of the S-S solutions and the effect of 
dissipation are also discussed. 
Our study is summarized in Section \ref{summary}.
\section{Formulation}\label{formulation}
\subsection{Basic equations}
In this paper, we consider S-S solutions describing 
runaway condensation at a region where the cooling rate 
dominates the heating rate.
The S-S solutions describe the evolution of the fluid 
after its memories of initial and 
boundary conditions have been lost during the runaway condensation. 
Therefore, the solutions are independent of how a cooling-dominated 
region forms. 

A net cooling rate $\rho{\cal L}=\Lambda - \Gamma$ is considered, where
$\Lambda$ and $\Gamma$ are the cooling and heating rates per unit volume 
and time, respectively.
In low-density regions, 
the main cooling process is spontaneous emission by collisionally 
excited gases.
In this situation, the cooling rate is determined by 
the collision rate, which is proportional to $\rho^2$.
Therefore, we adopt the following formula as 
the net cooling rate:
\begin{equation}
\rho{\cal L}(\rho,c_\mathrm{s}) 
= \Lambda_0 \rho^2 c_\mathrm{s}^{2\alpha}\;\;\mathrm{erg\;cm^{-3}\;s^{-1}}
,\;\;c_\mathrm{s}=\sqrt{\gamma P/\rho},
\label{cool rate}
\end{equation}
where $P$, $c_\mathrm{s}$, $\gamma$ and $\alpha$ are the pressure, 
sound speed,  
the ratio of specific heat, and a free parameter, respectively.
The effects of viscosity and thermal conduction are neglected for simplicity.

The basic equations for a radiative gas under a plane-parallel geometry are
the continuity equation,
\begin{equation}
\frac{\partial \rho}{\partial t} + \frac{\partial}{\partial x}(\rho v)=0,
\label{eoc}
\end{equation}
the equation of motion,
\begin{equation}
\frac{\mathrm{D} v}{\mathrm{D} t}
+\frac{1}{\gamma\rho}\frac{\partial }{\partial x}(c_\mathrm{s}^2\rho)=0,
\label{eom}
\end{equation}
and the energy equation,
\begin{equation}
\frac{2}{\gamma-1}\frac{\mathrm{D} \ln c_\mathrm{s}}{\mathrm{D} t} 
- \frac{\mathrm{D} \ln\rho}{\mathrm{D} t}  
= -\gamma\Lambda_0\rho c_\mathrm{s}^{2(\alpha-1)},
\label{eoe}
\end{equation}
where $\mathrm{D}/\mathrm{D}t=\partial/\partial t + v\partial/
\partial x$ indicates Lagrangian time derivative.
\subsection{Derivation of self-similar equations}
The density increases infinitely
because the net cooling rate is assumed to be 
$\propto \rho^2 c_\mathrm{s}^{2\alpha}$
in equation (\ref{cool rate}). 
The epoch at which the density becomes infinite is defined by $t_\mathrm{c}$.
We introduce a similarity variable $\xi$ and assume 
that the each physical quantity is given by the following form:
\[
x = x_0(t)\xi,\;\;v(x,t) = v_0(t)V(\xi),\;\;c_\mathrm{s}(x,t) = c_0(t)X(\xi),
\]
\begin{equation}
\rho(x,t) = \rho_0(t)\Omega(\xi)\;\;\mathrm{and}\;\;P(x,t)=P_0(t)\Pi(\xi).
\label{self vari1}
\end{equation}
By assuming power-law time dependence of $x_0(t)=at_*^n$, 
where $t_*=1-t/t_\mathrm{c}$ 
and $n=1/(1-\omega)$, we find the following relations:
\[
x_0(t) = at_*^n,\;\;v_0(t) = c_0(t) = \frac{na}{t_\mathrm{c}}t_*^{n\omega},\;\;
\]
\begin{equation}
\rho_0(t) = \left(\frac{n}{t_\mathrm{c}}\right)^{-2\alpha+3}
\frac{a^{2(1-\alpha)}}{\Lambda_0}t_*^{n\beta},\;\;\mathrm{and}\;\; 
P_0(t) = \frac{\rho_0c_0^2}{\gamma},
\label{self vari2}
\end{equation}
where $\beta=\omega(3-2\alpha)-1$.
Because the dimensional scale is introduced only by $\Lambda_0$, 
the time dependence of the physical quantities cannot be fixed.
Therefore, we leave $\omega$ as a free parameter. 

For convenience, instead of $\omega$, we 
can use a parameter $\eta$ which is given by
\begin{equation}
\eta= - \frac{(2-\alpha)\{(3-2\alpha)\omega-1\}}{1-\omega}.
\end{equation}
Using $\eta$, the time dependence of $\rho_0(t)$ and $P_0(t)$ is given by
\begin{equation}
\rho_0(t) \propto t_*^{-\eta/(2-\alpha)}\;\;\mathrm{and}\;\;
P_0(t) \propto t_*^{(1-\eta)/(1-\alpha)}.
\label{den P time}
\end{equation}
Substituting equations (\ref{self vari1}) and (\ref{self vari2})
into the basic equations (\ref{eoc}), (\ref{eom}) and (\ref{eoe}),
the similarity equations are obtained as
\begin{equation}
(V + \xi)\frac{\mathrm{d} \ln \Omega}{\mathrm{d} \xi} 
+ \frac{\mathrm{d} V}{\mathrm{d} \xi} = \beta,
\label{sim1}
\end{equation}
\begin{equation}
(V + \xi)\frac{\mathrm{d} V}{\mathrm{d} \xi} 
+\frac{X^2}{\gamma}\frac{\mathrm{d} \ln \Omega X^2}{\mathrm{d} \xi} = \omega V,
\label{sim2}
\end{equation}
and
\begin{eqnarray}
        &&\hspace{-0.7cm}(V + \xi)\left[\frac{2}{\gamma-1}\frac{\mathrm{d} \ln X}{\mathrm{d} \xi} 
- \frac{\mathrm{d} \ln \Omega}
{\mathrm{d} \xi}\right] \nonumber\\ 
&&\hspace{3cm}= \frac{2}{\gamma-1}\omega -\beta -\gamma \Omega
X^{2(\alpha-1)}.
\label{sim3}
\end{eqnarray}
\subsection{Homologous special solutions}\label{homo sol}
The similarity equations (\ref{sim1}), (\ref{sim2}) and (\ref{sim3}) 
have homologous special solutions with 
a spatially uniform density and sound speed.
Substituting $\Omega(\xi)=\Omega_0$ and $X(\xi)=X_0$ 
into the similarity equations
(\ref{sim1}), (\ref{sim2}) and (\ref{sim3}), one obtains
\begin{equation}
\frac{\mathrm{d}V}{\mathrm{d}\xi}=\beta,
\label{homo1}
\end{equation}
\begin{equation}
(V+\xi)\frac{\mathrm{d}V}{\mathrm{d}\xi}=\omega V,
\label{homo2}
\end{equation}
and 
\begin{equation}
\frac{2\omega}{\gamma-1}-\beta-\gamma\Omega_0 X_0^{2(\alpha-1)}=0.
\label{homo3}
\end{equation}
Equation (\ref{homo1}) is integrated to give
\begin{equation}
V=\beta\xi,
\label{v homo}
\end{equation}
where $V(\xi=0)=0$ is assumed.
Substituting equation (\ref{v homo}), equation (\ref{homo2}) becomes
\begin{equation}
\beta\omega(\alpha - 1)=0.
\end{equation}
Therefore, homologous solutions exist for 
$\beta=0$, $\omega=0$ or $\alpha=1$.
Equation (\ref{homo3}) gives the relation between $\Omega_0$ and $X_0$.
For $\beta=0$, the solution is given by $\Omega(\xi)=\Omega_0$, $V(\xi)=0$ and 
$X(\xi)=X_0$.
The time dependences of density and sound speed are given by
\begin{equation}
\rho(t)=\mathrm{const}.\;\;\mathrm{and}\;\;\;
c_\mathrm{s}(t) \propto t_*^{1/(2-2\alpha)},
\end{equation}
respectively.  This is the homologous isochoric mode. 
For $\alpha\geq1$, the solution is unphysical because 
the sound speed is negative or complex.
Therefore, 
no homologous isochoric solutions exist for $\alpha\geq1$.
This fact is related to the Balbus criterion 
for the isochoric instability \citep{B86}.
The solution with $\omega=0$ indicates isothermal condensation.
For $\alpha=1$, no S-S solutions are found except for the homologous 
solutions as shown in 
Section \ref{result}.
\subsection{Asymptotic behaviour}\label{asymp}
In this subsection, the asymptotic behaviour of the S-S solutions at 
$\xi\rightarrow\infty$ and $\xi\rightarrow0$ is investigated.
\subsubsection{Solutions for $\xi\rightarrow\infty$}
Assuming that $\Omega(\xi)$ is a decreasing function of $\xi$
and that 
$|V(\xi)|$ and $X(\xi)$ are increasing functions,
equations (\ref{sim1}), (\ref{sim2}) and (\ref{sim3}) become
\begin{equation}
\xi\frac{\mathrm{d}\ln \Omega}{\mathrm{d}\xi} \simeq \beta,\;\;
\xi\frac{\mathrm{d} V}{\mathrm{d}\xi} \simeq \omega V,\;\;\mathrm{and}\;\;
\xi\frac{\mathrm{d} \ln X}{\mathrm{d}\xi} \simeq \omega
\end{equation}
to the lowest order.
Expanding to the next order, the asymptotic solutions are given by
\begin{equation}
\Omega(\xi) \simeq \Omega_\infty \xi^\beta
+ \Omega_\infty^{(1)}\xi^{\beta+\omega-1},
\end{equation}
\begin{equation}
V(\xi)\simeq V_\infty\xi^\omega +
V_\infty^{(1)}\xi^{2\omega-1},
\end{equation}
and 
\begin{equation}
X(\xi)  \simeq X_\infty\xi^\omega
+X_\infty^{(1)}\xi^{2\omega-1},
\end{equation}
where
\begin{equation}
\Omega_\infty^{(1)} = \Omega_\infty V_\infty\frac{\beta+\omega}{1-\omega},
\end{equation}
\begin{equation}
V_\infty^{(1)}=\frac{1}{1-\omega}
\left[\omega V_\infty^2 + \frac{X_\infty^2}{\gamma}(\beta+2\omega)\right],
\end{equation}
and 
\begin{equation}
X_\infty^{(1)}=\frac{(\gamma+1)\omega V_\infty X_\infty
+\gamma(\gamma-1)\Omega_\infty X_\infty^{2(\alpha-1)}}{2(1-\omega)}.
\end{equation}
\subsubsection{Solutions for $\xi\rightarrow 0$}
We expand $\Omega(\xi)$, $V(\xi)$ and $X(\xi)$ in the following forms:
\[
\Omega(\xi) \simeq \Omega_{00} + \Omega_{00}^{(1)}\xi^\nu,\;\;
V(\xi) \simeq V_{00}\xi + V_{00}^{(1)}\xi^{\nu + 1},
\]
and 
\begin{equation}
X(\xi) \simeq X_{00} +X_{00}^{(1)}\xi^\nu.
\label{asym 0}
\end{equation}
Substituting equations (\ref{asym 0}) into (\ref{sim1}), 
(\ref{sim2}) and (\ref{sim3}),
we obtain the following relations for coefficients: 
\begin{equation}
        V_{00} = \beta\;\;\mathrm{and}\;\;
V_{00}^{(1)}= - \frac{\nu(\beta+1)\Omega_{00}^{(1)}}{(\nu+1)\Omega_{00}}.
\end{equation}
The relation between $\Omega_{00}$ and $X_{00}$ is given by 
\begin{equation}
\gamma(\gamma-1)\Omega_{00}X_{00}^{2(\alpha-1)} = 
2\omega-\beta(\gamma-1).
\label{asym1 0}
\end{equation}
The relation between $\Omega_{00}^{(1)}$ and $X_{00}^{(1)}$ is given by 
\begin{equation}
\frac{\Omega_{00}^{(1)}}{\Omega_{00}}+2\frac{X_{00}^{(1)}}{X_{00}}=0,
\end{equation}
and $\nu$ is given by 
\begin{equation}
\nu=\frac{(2-\alpha)\epsilon_{00}}{\beta+1}
,\;\;\mathrm{where}\;\;\epsilon_{00}=(\gamma-1)\Omega_{00}X_{00}^{2(\alpha-1)}.
\end{equation}
\subsection{Critical point}\label{sonic point}
Equations (\ref{sim1}), (\ref{sim2}) and (\ref{sim3}) can be rewritten as
\begin{equation}
\frac{\mathrm{d} \ln \Omega}{\mathrm{d} \xi} = \frac{I_2}{I_1},\hspace{5mm}
\frac{\mathrm{d} V}{\mathrm{d} \xi} = \frac{I_3}{I_1}\;\;\;\mathrm{and}\;\;\;
\frac{\mathrm{d} \ln X}{\mathrm{d} \xi} = \frac{I_4}{I_1},
\label{num eq}
\end{equation}
where
\begin{equation}
I_1 = (V + \xi)^2 - X^2,\;\;\;\:
I_2= \frac{X^2}{V+\xi}g + f,
\end{equation}
\begin{equation}
I_3 = -X^2 g +\beta I_1 -(V+\xi)f,
\end{equation}
\begin{equation}
I_4 = -\frac{1}{2}\left[\frac{X^2 - \gamma(V+\xi)^2}
{V+\xi}\right]g + \frac{\gamma-1}{2}f,
\end{equation}
\begin{equation}
g(\Omega,X) = 
(\gamma-1)\left\{\Omega_{00} X_{00}^{2(\alpha-1)}- \Omega X^{2(\alpha-1)}\right\},
\end{equation}
and 
\begin{equation}
f(\xi,V) = \beta(V+\xi) - \omega V.
\end{equation}
A singular point exists where $I_1=0$ 
in each of the equations (\ref{num eq}). This
is the critical point ($\xi=\xi_\mathrm{s}$).
In the S-S solutions which is smooth at this point, the numerators 
($I_2$, $I_3$ and $I_4$) must vanish at $\xi=\xi_\mathrm{s}$.
Among four conditions $I_1=I_2=I_3=I_4=0$, there are two independent 
conditions which are given by
\begin{equation}
(V + \xi)^2 = X^2 \hspace{0.5cm}\mathrm{and}\hspace{0.5cm}
\frac{X^2}{V+\xi}g + f = 0.
\label{sonic1}
\end{equation}
\subsubsection{Topological property of the critical point}
In this subsection, the topological properties of the critical point are investigated.
We introduce a new variable $s$ which is defined by 
$\mathrm{d}s = I_1\mathrm{d}\xi$
\citep{WS85}.
The dimensionless quantities at infinitesimal distances 
from the critical point, 
$s_\mathrm{s} + \delta s$, are given by
\begin{equation}
\mbox{\boldmath{$Q$}}(s_\mathrm{s}+\delta s) 
= \mbox{\boldmath{$Q_\mathrm{s}$}}  + \mbox{\boldmath{$\delta Q$}},
\label{topo}
\end{equation}
where $\mbox{\boldmath{$Q$}}=(\xi,\ln \Omega,V,\ln X)$. The subscript "$s$" 
indicates the value at the critical point.
Substituting equation (\ref{topo}) into equations 
(\ref{num eq}) and linearizing, one obtains
\begin{equation}
\frac{\mathrm{d} \delta Q_i}{\mathrm{d}s}
=A_{ij}|_{\xi=\xi_\mathrm{s}} \delta Q_j,\;\;\;\;A_{ij}=\frac{\partial I_i}{\partial Q_j}.
\label{sonic grad}
\end{equation}
The eigenvalues $\lambda$ of $A_{ij}$ provide the topological property of the critical point.
The eigenequation is given by
\begin{equation}
\left(\frac{\lambda}{X_\mathrm{s}}\right)^2\left[
\left(\frac{\lambda}{X_\mathrm{s}}\right)^2 + b_1 \frac{\lambda}{X_\mathrm{s}} + b_2\right]=0,
\label{eigen eq}
\end{equation}
where
\begin{equation}
b_1= \epsilon_\mathrm{s}\left\{(\gamma-1)(\alpha-1)+1\right\} + \omega + \beta - 2,
\end{equation}
and
\begin{eqnarray}
b_2&=&2\gamma(\alpha-1)\epsilon_\mathrm{s}^2 \nonumber \\
&+&(-\beta\gamma-4\omega\alpha+2\omega-2\gamma\alpha
+2\alpha-3-2\beta+\omega\gamma+\gamma)\epsilon_\mathrm{s} \nonumber \\
&+&4\omega-\beta+2\omega^2-\frac{4\omega^2-\beta-2\omega+2\omega\beta}{\gamma},
\end{eqnarray}
where $\epsilon_\mathrm{s} = (\gamma-1)\Omega_\mathrm{s} X_\mathrm{s}^{2(\alpha-1)}$. 
The degenerate solution, $\lambda=0$, is attributed to the homologous 
special solutions (see Subsection \ref{homo sol}). 
Therefore, the topology is described by the two eigenvalues 
other than $\lambda=0$.
\subsection{Numerical methods}\label{numerical}
In this subsection, our numerical method for solving the similarity equations
(\ref{num eq}) is described.
We adopt $s$ as an integrating variable. 
Using $s$, equations (\ref{num eq})
are rewritten as 
\begin{equation}
        \frac{\mathrm{d} Q_i}{\mathrm{d} s} = I_i,\;\;\; i=1,2,3\;\;\mathrm{and}\;\;4,
\label{num eq1}
\end{equation}
which are not singular at the critical point.

The critical point is specified by ($\xi_\mathrm{s}$, $V_\mathrm{s}$)
because there are four unknown quantities 
$\mbox{\boldmath{$Q_\mathrm{s}$}}$ and 
two conditions, namely equations (\ref{sonic1}). Moreover, the similarity 
equations (\ref{num eq1}) 
are invariant under the transformation 
$\xi\rightarrow m\xi$, $V\rightarrow mV$, $\Omega\rightarrow 
m^{-2(\alpha-1)}\Omega$ and $X\rightarrow mX$, 
where $m$ is an arbitrary constant.
Therefore, the S-S solution for $(m\xi_\mathrm{s},mV_\mathrm{s})$
is the same as that for $(\xi_\mathrm{s},V_\mathrm{s})$.
As one can take $m$ in such a way that $(\xi_\mathrm{s},V_\mathrm{s})$
satisfies the relation $\xi_\mathrm{s}^2+V_\mathrm{s}^2=1$,
the critical point is specified only by $\xi_\mathrm{s}$.

Given $\xi_\mathrm{s}$, the velocity is obtained as 
$V_\mathrm{s}=-\sqrt{1-\xi_\mathrm{s}^2}$.
From equations (\ref{sonic1}), $\Omega_\mathrm{s}$ 
and $X_\mathrm{s}$ are determined.
As $X_\mathrm{s}=V_\mathrm{s}+\xi_\mathrm{s}$ is positive, 
the range of $\xi_\mathrm{s}$ is $1/\sqrt{2}<\xi_\mathrm{s}<1$.
The similarity equations (\ref{num eq1}) are integrated 
from $\xi_\mathrm{s}$ to both directions 
($\xi\rightarrow0$ and $\xi\rightarrow \infty$) along the 
gradient derived from equation 
(\ref{sonic grad}) using the fourth-order Runge-Kutta method.
The position of the critical point $\xi_\mathrm{s}$ is determined 
iteratively by the bisection method 
until the solution satisfies the asymptotic behaviours (see Subsection \ref{asymp}).
\section{Results}\label{result}
\subsection{Physically possible range of parameters $(\eta,\alpha)$}
In this section, we constrain the parameter space $(\eta,\alpha)$ by 
physical properties of the flow.
\begin{figure}
 \centering
     \includegraphics[width=90mm]{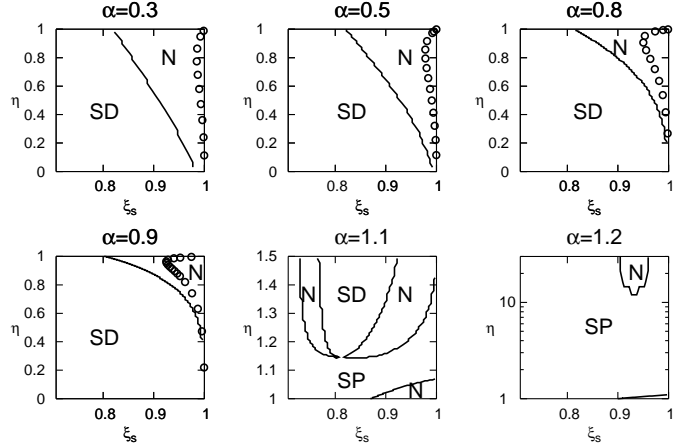}
 \caption{Topological properties of the critical point 
 for $\alpha=$0.3, 0.5, 0.8, 0.9, 1.1 and 1.2. 
 The ordinates and abscissas are $\eta$ and $\xi_\mathrm{s}$, respectively.
 The letters `N', `SD' and `SP' indicate a node, a saddle and a spiral, 
 respectively. The open circles indicate S-S solutions.
 }\label{a0.5mach}
\label{sonic}
\end{figure}
\subsubsection{Range of $\eta$}\label{eta range}
Using equation (\ref{den P time}),
the time dependences of the central density and pressure 
are given by 
\begin{equation}
\rho_{00}(t) \propto t_*^{-\eta/(2-\alpha)}\;\;\mathrm{and}\;\;
P_{00}(t) \propto t_*^{(1-\eta)/(1-\alpha)},
\label{den P 00}
\end{equation}
where the subscript "00" indicates the value at the centre.
During S-S condensation by cooling, the central density 
(pressure) must increases 
(decreases) monotonically with time.
Therefore, the following two conditions are obtained:
\begin{equation}
-\eta/(2-\alpha)<0\;\;\;\Rightarrow \eta>0
\label{region1}
\end{equation}
and 
\begin{eqnarray}
(1-\eta)/(1-\alpha)>0\;\;\;\Rightarrow 
\left\{
\begin{array}{cccc}
\eta<1 & \mathrm{for} & \alpha<1 \\
\eta>1 & \mathrm{for} & \alpha>1 & \hspace{-2mm}.\\
\end{array}
\right.
\label{region2}
\end{eqnarray}
From equations (\ref{region1}) and (\ref{region2}), 
the range of $\eta$ is given by
\begin{eqnarray}
\left\{
\begin{array}{cccc}
0<\eta<1 & \mathrm{for} & \alpha <1 &\\ 
\eta>1 & \mathrm{for} & \alpha >1 &\hspace{-2mm}. \\ 
\end{array}
\right.
\label{omega range}
\end{eqnarray}
For $\eta \sim 0$, the time dependence of $\rho_{00}$ and $P_{00}$ is
given by
\begin{equation}
\rho_{00}(t) \simeq \mathrm{const.} \;\;\;\mathrm{and}\;\;\;
P_{00}(t)\propto t_*^{1/(1-\alpha)}.
\end{equation}
This time evolution indicates the isochoric mode.
For $\eta \sim 1$, the time dependence of $\rho_{00}$ and $P_{00}$ is
given by
\begin{equation}
\rho_{00}(t)\propto t_*^{-1/(2-\alpha)}\;\;\;\mathrm{and}\;\;\;
P_{00}(t)\simeq \mathrm{const.},
\end{equation}
corresponding to the isobaric mode.
The critical value of $\eta=\eta_\mathrm{eq}$ is given by the condition that 
the increasing rate of $\rho_{00}(t)$ is 
equal to the rate of decrease of $P_{00}(t)$, which is 
given by 
\begin{equation}
\eta_\mathrm{eq}=(2-\alpha)/(3-2\alpha).
\end{equation}
For $\eta=\eta_\mathrm{eq}$, the time dependences of $\rho_{00}$ and $P_{00}$ 
are given by
\begin{equation}
\rho_{00}(t)\propto t_*^{-1/(3-2\alpha)}\;\;\;\mathrm{and}\;\;\;
P_{00}(t)\propto t_*^{1/(3-2\alpha)}.
\end{equation}
Therefore, the S-S solutions for $0<\eta<\eta_\mathrm{eq}$ and 
$\eta_\mathrm{eq}<\eta<1$
are close to the isochoric and the isobaric modes, respectively.
\begin{figure*}
 \centering
     \includegraphics[width=110mm]{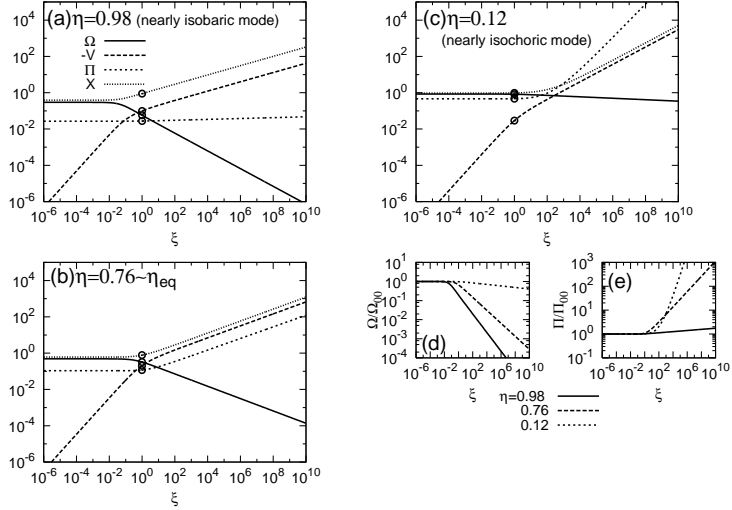}
 \caption{The distribution of dimensionless quantities 
          ($\Omega$,$-V$,$\Pi$,$X$) 
          for (a)$\eta$=0.98, (b)0.76 and (c)0.12. 
          The open circles indicate the critical points.
          Panels (d) and (e) indicate the dependence of $\Omega/\Omega_{00}$ 
          and $\Pi/\Pi_{00}$, respectively, on $\eta$.
 }
\label{ome}
\end{figure*}
\begin{figure*}
 \centering
     \includegraphics[width=110mm]{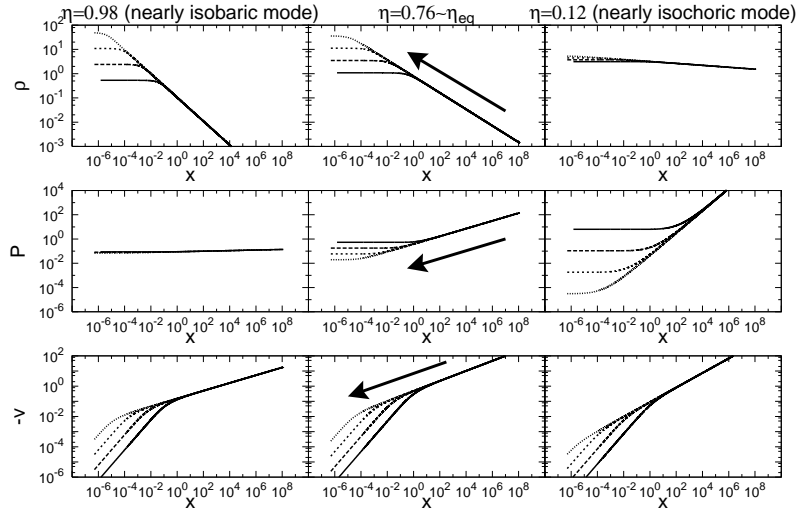}
 \caption{Time evolution of dimensional quantities ($\rho$, $P$, $-v$).
         The solid, long-dashed, short-dashed and dotted lines are 
        corresponding to 
        $t_*$=1, 0.1, 0.01 and 0.001, respectively. The columns 
         correspond to 
         $\eta$=0.98, 0.76 and 0.12 from left to right, respectively.
         The arrows indicate the time evolution of flow.
         } 
 \label{ome time evo}
\end{figure*}
\subsubsection{The range of $\alpha$}
In the similarity equations (\ref{num eq1}),
the critical point is specified by $\xi_\mathrm{s}$ 
for given $\alpha$ and $\eta$.
The topology of the critical point is given by the two eigenvalues of 
the eigenequation (\ref{eigen eq}) other than $\lambda=0$.
When the two eigenvalues are real and have the same sign, 
the critical point is a node.
When the two eigenvalues are real and have the opposite signs, 
the critical point is a saddle.
When the two eigenvalues are complex, the critical point is a spiral.
Fig. \ref{sonic} shows the topology of the critical 
point in the parameter space 
($\xi_\mathrm{s}$, $\eta$) for $\alpha=$0.3, 0.5, 0.8, 0.9, 1.1 and 1.2. 
In Fig. \ref{sonic}, the letters `N', `SD' and `SP' indicate a node, 
a saddle and a spiral, respectively.
The open circles indicate the numerically obtained S-S solutions presented 
in Subsection \ref{s-s property}.
Fig. \ref{sonic} shows that the topology drastically 
changes at $\alpha=1$.
For $\alpha<1$, because the critical point is a node or a saddle, 
the S-S solutions are expected to exist.
For $\alpha\simeq 1.1$, a non-negligible fraction of 
the parameter space is covered by a spiral.
Almost all parameter region is a spiral for $\alpha\ga1.2$.
We searched the S-S solutions that connect $0<\xi<\infty$ and found them only 
in N and SD for $\alpha< 1$.
\subsection{Self-similar solutions}\label{s-s property}
In this subsection, we describe new S-S solutions and investigate 
their properties. As shown in Subsection \ref{eta range}, 
our S-S solutions exist 
for $0<\eta<1$ and $\alpha<1$. 

Figs \ref{ome}(a), (b) and (c) show the obtained S-S solutions
for $\alpha=0.5$ with $\eta$=0.98, 0.76 and 0.12, respectively.
The ordinates denote the dimensionless quantities 
($\Omega$, $-V$, $\Pi$, $X$). The abscissas denote 
the dimensionless coordinate $\xi$.
The S-S solutions for $\eta$=0.98 and 0.12 approximately represent 
the isobaric and the isochoric modes, respectively.
The solution for $\eta=0.76\sim \eta_\mathrm{eq}$ 
corresponds to the intermediate mode shown in Subsection \ref{eta range}.
The open circles indicate the critical points.
In Figs \ref{ome}(a), (b) and (c), one can see that the S-S solutions 
depend strongly on $\eta$.
Let us focus on the dependence on $\eta$ of density 
and pressure, shown in Figs \ref{ome}(d) and (e), respectively. 
In these figures, it can be seen that, for $\eta=0.98$ (the isobaric mode), 
the density increases sharply towards $\xi=0$, whereas the 
pressure is almost spatially constant.
In contrast, for $\eta=0.12$ (the isochoric mode), 
the density is almost spatially 
constant whereas the pressure decreases sharply towards $\xi=0$. 
\begin{table*}
\begin{center}
\begin{tabular}{|l||c|c|c|c|c|c|c|c|c|}
\hline
$\eta$& $n$ & $-\eta/(2-\alpha)$ & $(1-\eta)/(1-\alpha)$ & $\xi_\mathrm{s}$ 
& $\Omega_{00}$ & $X_{00}$ & $\Omega_\infty$ 
   & $V_\infty$ & $X_\infty$  \\ \hline \hline 
0.059 & 1.923 &-0.039 & 1.882 &  0.9999 &0.873 & 0.977 & 0.882  & -0.0234 & 0.0408 \\ 
0.272 & 1.818 &-0.181 & 1.455 &  0.9975 &0.768 & 0.883 & 0.782  & -0.124  & 0.201 \\ 
0.500 & 1.667 &-0.333 & 1.000 &  0.9907 &0.643 & 0.766 & 0.612  & -0.244  & 0.391 \\ 
0.761 & 1.493 &-0.507 & 0.615 &  0.9792 &0.486 & 0.609 & 0.343  & -0.333  & 0.613 \\ 
0.945 & 1.370 &-0.631 & 0.110 &  0.9853 &0.360 & 0.472 & 0.123  & -0.223  & 0.805 \\ 
0.995 & 1.337 &-0.663 & 0.011 &  0.9986 &0.237 & 0.315 & 0.0260 & -0.0576 & 0.952 \\ 
\hline
\end{tabular}
\end{center}
\caption{Relevant parameters for obtained self-similar solutions ($\alpha=0.5$) 
are summarized. 
}
\label{a0.5 table}
\end{table*}

Next, the S-S solutions are shown using the dimensional physical quantities 
for various $\eta$.
Fig. \ref{ome time evo} shows the time evolution of 
$\rho$ (first row), $P$ (second row) and $-v$ (third row).
The columns in Fig. \ref{ome time evo} corresponds to 
$\eta$=0.98, 0.76 and 0.12 from left to right.
The solid, long-dashed, short-dashed and dotted lines correspond to
$t_*$=1, 0.1, 0.01 and 0.001, respectively.
From Fig. \ref{ome time evo}, one can clearly be seen 
that $\eta$ specifies how the gas condenses. 
In the case with $\eta$ = 0.98 (the isobaric mode), 
runaway condensation occurs and 
the density increases in the central region, 
whereas the pressure is spatially and temporally 
constant. 
In the case with $\eta$ = 0.12 (the isochoric mode), 
the density is spatially and temporally constant and
$P_{00}$ decreases considerably. 
In the case with $\eta$=0.76, 
the condensation occurs in an intermediate manner between 
the isobaric and the isochoric modes.

Relevant parameters, ($\eta$, $n$, $-\eta/(2-\alpha)$, $(1-\eta)/(1-\alpha)$, 
$\xi_\mathrm{s}$, $\Omega_{00}$, $X_{00}$, 
$\Omega_\infty$, $V_\infty$, $X_\infty$),
for the S-S solutions obtained for $\alpha=0.5$ are summarized 
in Table \ref{a0.5 table}.

\begin{figure}
 \centering
 \includegraphics[width=90mm]{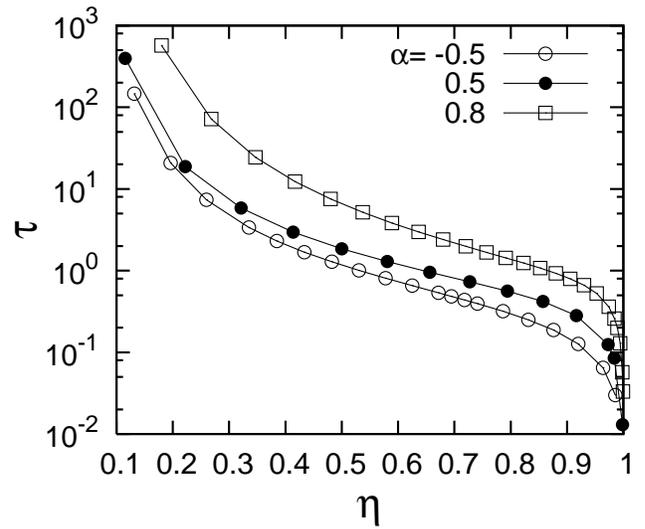}
 \caption{The dependence of $\tau$ (equation \ref{tau}) on $\eta$. 
          The open circles, filled circles and open boxes indicate 
          the S-S solutions for $\alpha$=-0.5, 0.5 and 0.8, respectively.}
\label{00}
\end{figure}

From the above results, it is confirmed that 
the S-S solutions are specified by $\eta$ 
and contain two asymptotic limits (the isobaric and the isochoric modes).
Moreover, the parameter $\eta$ is related to the ratio of the sound-crossing to
the cooling time-scales in the central region. Both time-scales are defined by 
\begin{equation}
t_\mathrm{sound}^\mathrm{cen}=\frac{x_\mathrm{cen}(t)}{c_{00}(t)}
=\frac{\xi_\mathrm{cen}}{n X_{00}}t_\mathrm{c}t_*
\label{tsound}
\end{equation}
and 
\begin{equation}
t_\mathrm{cool}^\mathrm{cen}
=\frac{P_{00}(t)}{(\gamma-1)\Lambda_0\rho_{00}(t)^2 c_{00}(t)^{2\alpha}}
=\frac{\Omega_{00}^{-1}X_{00}^{2(1-\alpha)}}{\gamma(\gamma-1)n} t_\mathrm{c}t_*,
\label{tcool}
\end{equation}
respectively, where $c_{00}$ is the sound speed at the centre, 
and the scale length $x_\mathrm{cen}(t)$ 
and $\xi_\mathrm{cen}$ are defined by 
\begin{equation}
\rho(x_\mathrm{cen}(t),t)/\rho(0,t)=\Omega(\xi_\mathrm{cen})/\Omega_{00}=0.8.
\end{equation}
From equations (\ref{tsound}) and (\ref{tcool}), 
the ratio $\tau$ of the above two time-scales is obtained as
\begin{equation}
\tau=\frac{t_\mathrm{sound}^\mathrm{cen}}{t_\mathrm{cool}^\mathrm{cen}}
= \xi_\mathrm{cen}\gamma(\gamma-1)\Omega_{00}X_{00}^{2\alpha-3}.
\label{tau}
\end{equation}
Fig. \ref{00} shows the dependence of $\tau$ on $\eta$ for 
$\alpha$=-0.5, 0.5 and 0.8.
The figure shows that $\tau\rightarrow0$ for $\eta\rightarrow1$ 
and $\tau\rightarrow\infty$ for $\eta \rightarrow0$ 
independent of $\alpha$.
For $\eta\sim1$, because $\tau\ll 1$,
the gas condenses in pressure equilibrium with its surroundings.
For $\eta\sim0$, 
because $\tau\gg1$,
the pressure decreases and the density increases only slightly.
These facts are consistent with the discussion in Subsection \ref{eta range} 
and with Fig. \ref{ome time evo}.

Finally, we consider the range of $\alpha$ for which the S-S solutions exist.
When $\alpha=1.0$, no S-S solutions are found except for the homologous 
special solutions (see Subsection \ref{homo sol}).
Moreover, as mentioned in Subsection \ref{homo sol}, 
there are no homologous isochoric solutions
for $\alpha>1$.
Therefore, for $\alpha>1$, the condensation of the fluid is not expected to 
be self-similar.
\citet{F07} found a family of exact solutions for $\alpha=1.5>1$.
Their solutions are not self-similar because 
$t_\mathrm{sound}^\mathrm{cen}$ 
and $t_\mathrm{cool}^\mathrm{cen}$ in the central region
obey the following different scaling laws: 
$t_\mathrm{sound}^\mathrm{cen}\propto t_*^{3/2}$ 
and $t_\mathrm{cool}^\mathrm{cen}\propto t_*$.
In their solution, during condensation, 
as $t_\mathrm{sound}^\mathrm{cen}$ becomes much smaller 
than $t_\mathrm{cool}^\mathrm{cen}$, 
the pressure of the central region is expected to be constant and the isobaric 
approximation is locally valid.

\begin{figure*}
 \centering
     \includegraphics[width=110mm]{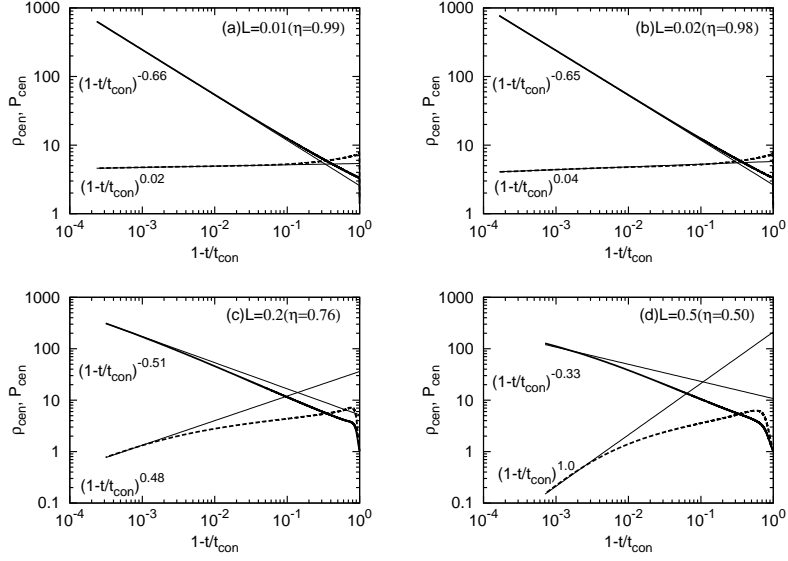}
     \caption{The time evolution of $\rho_{00}$ (thick lines) and $P_{00}$ 
     (dashed lines) as a function of $1-t/t_\mathrm{con}$
     for (a)$L$=0.01, (b)0.02, (c)0.2 and (d)0.5.
     The thin solid lines indicate the corresponding S-S solutions.}
\label{ve2 00}
\end{figure*}
\begin{figure*}
 \centering
     \includegraphics[width=110mm]{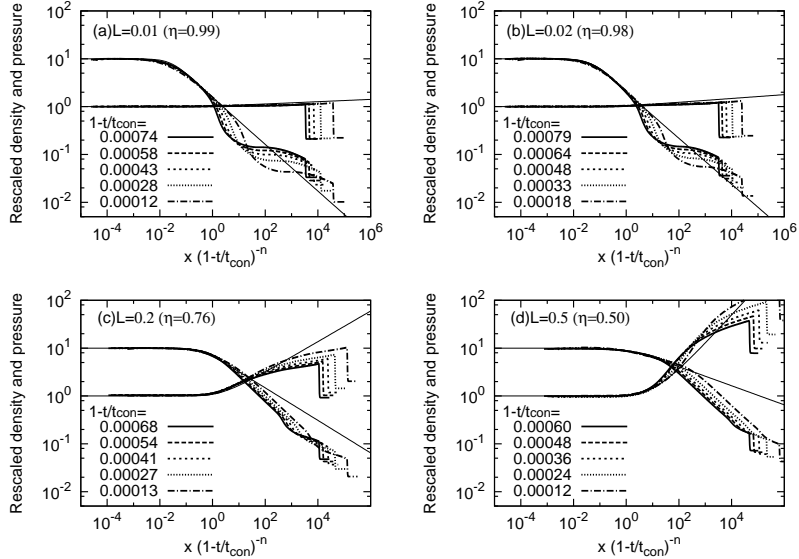}
     \caption{The distribution of the rescaled density 
     $\rho(1-t/t_\mathrm{con})^
     {\eta/(2-\alpha)}$ and pressure $P(1-t/t_\mathrm{con})^
     {-(1-\eta)/(2-\alpha)}$
     as a function of scaled coordinate $x(1-t/t_\mathrm{con})^{-n}$.
     The $L$ and $\eta$ in each panel are the same as that in Fig. \ref{ve2 00}.
     The lines are labelled according to $1-t/t_\mathrm{con}$.
     The rescaled density and pressure at the centre are set to 10 and 0.1, 
     respectively.
     The thin solid lines indicate the corresponding S-S solutions.}
\label{ve2}
\end{figure*}
\section{Discussion}\label{discuss}
\subsection{Comparison with the results of time-dependent 
numerical hydrodynamics}
We compare the S-S solutions with 
results of time-dependent numerical hydrodynamics 
using one-dimensional Lagrangian 2nd-order Godunov method \citep{vL97}.
A converging flow of an initially uniform and thermal-equilibrium gas 
($\rho(x)=P(x)=1$), is considered. 
An initial velocity field is given by 
$v(x)= -2\tanh(x/L)$, where $L$ is a parameter.
The net cooling rate is assumed to be 
$\rho{\cal L}(\rho,T)=(\rho^{3/2}P^{1/2} - \rho)/(\gamma-1)$, where 
$\alpha$=0.5.
The inflow creates two shock waves, which propagate outward.
Runaway condensation of the perturbation owing to TI 
occurs in the post-shock region because the cooling 
rate dominates the heating rate. The central density continues to grow and 
ultimately becomes infinite. During the runaway condensation, 
the flow in the central region is expected to lose the memory of the 
initial and the boundary conditions,
and to converge to one of the S-S solutions.
The scale of perturbation in the post-shock region can be controlled by 
the value of $L$.
The larger $L$ is, the larger the scale of perturbation becomes.
We perform numerical simulations for the case with 
$L$=0.005, 0.01, 0.02, 0.2 and 0.5.

Fig. \ref{ve2 00} shows the time evolution of 
$\rho_\mathrm{cen}(t)$ and $P_\mathrm{cen}(t)$ as 
a function of $1-t/t_\mathrm{con}$,
where $\rho_\mathrm{cen}$ and $P_\mathrm{cen}$ are 
the central density and pressure,
and $t_\mathrm{con}$ is an epoch when $\rho_\mathrm{cen}$ becomes infinity 
estimated from the numerical results. 
The panels represent (a)$L$=0.01, (b)0.02, (c)0.2 and (d)0.5,
respectively. 
Using equation (\ref{den P 00}), 
the time evolution of $\rho_\mathrm{cen}(t)$ and $P_\mathrm{cen}(t)$ 
of each simulation provides the corresponding $\eta$.
As a result, the corresponding values of $\eta$ are found to be
(a)0.99, (b)0.98, (c)0.76 and (d)0.50.
The thin solid lines in Fig. \ref{ve2 00} indicate 
the corresponding S-S solutions.
From the value of $\eta$ and Fig. \ref{ve2 00}, 
cases (a) and (b) correspond to the isobaric mode, 
case (c) corresponds to 
the intermediate mode 
($\eta\sim\eta_\mathrm{eq}$) and case (d) corresponds to 
the isochoric mode.
There is a difference in the convergence speed from
the corresponding S-S solution. Fig. \ref{ve2 00} shows that 
the convergence speed of (a) and (b) is faster than that of (c) and (d).
This is because the scale-length of the condensed region of 
(a) and (b) is smaller 
and the sound-crossing time-scale is shorter.

Fig. \ref{ve2} shows the time evolution of the rescaled density 
$\rho(1-t/t_\mathrm{con})^{\eta/(2-\alpha)}$ and pressure 
$P(1-t/t_\mathrm{con})^{-(1-\eta)/(1-\alpha)}$ as a function of the 
rescaled coordinate 
$x(1-t/t_\mathrm{con})^{-n}$.
The parameters $L$ and $\eta$ in Fig. \ref{ve2} are the same 
as those in Fig. \ref{ve2 00}.
The thin solid lines in Fig. \ref{ve2} indicate the corresponding S-S solutions.
In all the cases (a-d), the central regions are well approximated by 
the corresponding S-S solutions.
From Fig. \ref{ve2}, it can be seen that the region that is well approximated 
by the S-S solution spreads with time in each panel.

\begin{figure}
 \centering
     \includegraphics[width=80mm]{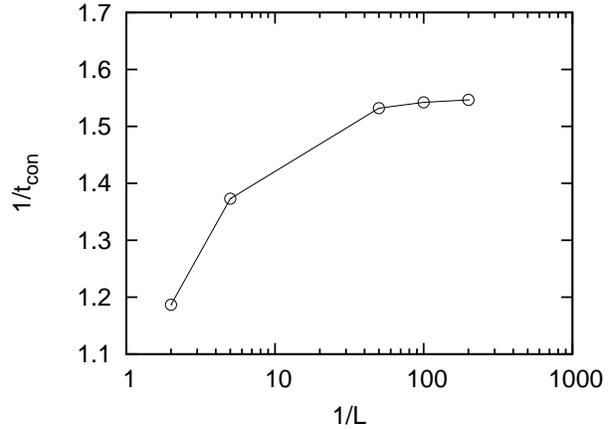}
     \caption{The dependence of the growth rate, $1/t_\mathrm{con}$, on 
     the wavenumber, $1/L$.
     }
\label{ve2 collapse}
\end{figure}

From the above discussion, the non-linear evolution of perturbation 
is well approximated by one of the S-S solution in high density limit.
Which S-S solution ($0<\eta<1$) is more likely to be realised in 
actual environments where various scale perturbations exist?
To answer this question, the dependence of 
$t_\mathrm{con}$ on $L$ is investigated.
Fig. \ref{ve2 collapse} shows the dependence of $1/t_\mathrm{con}$ on 
$1/L$. Here, $1/L$ and $1/t_\mathrm{con}$ correspond to 
the wavenumber and the growth rate of the perturbation, respectively.
Therefore, Fig. \ref{ve2 collapse} can be interpreted as an approximate 
dispersion relation for the non-linear evolution of TI.
For larger $1/L$, $1/t_\mathrm{con}$ is larger and 
$1/t_\mathrm{con}$ appears to approach a certain asymptotic value.
A similar behaviour is also found in the dispersion relation of \citet{F65} 
without heat conduction.
In Fig. \ref{ve2 collapse}, the condensation for $1/L=5.0$ corresponds to 
the intermediate mode. 
Furthermore, although the isobaric condensation grows faster,
there is little difference between $1/t_\mathrm{con}$
for $1/L$=200 (the isobaric mode) and 2 (the isochoric mode). 
Therefore, both the isobaric and the isochoric condensations are 
expected to coexist in actual astrophysical environments.
\subsection{Astrophysical implication}
In astrophysical environment, there are roughly two ranges of temperature 
with $\alpha<1$ where our S-S solutions are applicable.

In the high-temperature case, $T\ga 2\times 10^5$K
\citep{SD93}, the dominant coolants are metal lines and Bremsstrahlung.
In a shock with a velocity greater than 
\begin{equation}
        V\ga 120\;\mathrm{km\;s^{-1}}\left(\frac{T}{2\times10^5\;\mathrm{K}}\right)^{1/2},
\end{equation}
the post-shock region is thermally unstable ($\alpha<1$).
For example, a typical supernova satisfies 
this condition, and the S-S condensation is expected to occur.
In reality, the density does not become infinite because the cooling time-scale 
increases when the gas reach the thermally stable phase.
In this case, the cooled layer will rebound and overstability may take place
\citep{CI82,YN01,S03}.

In the low-temperature case, 300$\la T\la$6000K, 
the dominant coolant is neutral carbon atom.
\citet{KI00} suggested that 
the structure of the ISM is caused by TI
during the phase transition (WNM $\rightarrow$ CNM).
The phase transition is induced by a shock wave.
\citet{HP99} considered the converging flow of a WNM under 
a plane-parallel geometry 
using one-dimensional hydrodynamical calculations, and 
investigated the condensation condition of 
the Mach number and pressure in the pre-shock region.
In situations which satisfy their condition, S-S condensation is expected to 
occur. 
\subsection{Effects of dissipation}
In the actual gas, the effect of viscosity and heat conduction
becomes important on small scales. The importance of dissipation can be 
evaluated by the ratio between the advection and the dissipation terms, or 
Raynolds number, which is given by 
\begin{equation}
{\cal R}=
\frac{C_\mathrm{V} \rho v\partial_x T}{\partial_x(K(T)\partial_x T)} \sim 
\frac{C_\mathrm{V}\rho_{00} U x_\mathrm{cen}}{K(T_{00})},
\label{raynolds}
\end{equation}
where $C_\mathrm{V}$, $K(T)$ and $U$ are the specific heat at constant volume, 
the heat conduction coefficient and the typical velocity, respectively. 
The Plandtl number is implicitly assumed to be of order unity.
A typical length-scale is assumed to be $x_\mathrm{cen}(t)$.
If the flow converges to one of the S-S solutions, the ratio, $\tau$, 
between the sound-crossing and the cooling timescales is constant.
Therefore $\tau$ is defined by
\begin{equation}
\tau = \frac{t_\mathrm{sound}^\mathrm{cen}}{t_\mathrm{cool}^\mathrm{cen}} 
= x_\mathrm{cen}\gamma (\gamma-1)\Lambda_0
\rho_{00} c_{00}^{2\alpha-3}.
\end{equation}
Using $\tau$, $x_\mathrm{cen}$ is written as
\begin{equation}
x_\mathrm{cen}= \frac{\tau}{\gamma (\gamma-1)\Lambda_0}
\rho_{00}^{-1} c_{00}^{3-2\alpha}.
\end{equation}
Substituting this equation into equation (\ref{raynolds}), one obtains
\begin{equation}
{\cal R}\sim\frac{C_\mathrm{V}f_v \tau}{\gamma
(\gamma-1)\Lambda_0K_0}c_{00}^{4-2\alpha-\kappa},
\label{ray}
\end{equation}
where $U=f_v c_{00}$$(f_v<1)$ and $K(T_{00})=K_0c_{00}^\kappa$.

For the low-temperature case, 
we adopt 
$K = 2.5\times10^3 \sqrt{T}\;\;$ ergs cm$^{-1}$ K$^{-1}$ s$^{-1}$ \citep{P53} 
and the following cooling function \citep{KI02}:
\begin{equation}
\Lambda(T) \sim 1.0\times10^{20} \rho^2\sqrt{T}\;\;\;\mathrm{erg\;cm^{-3}\;s^{-1}}.
\end{equation}
From equation (\ref{ray}), the Raynolds number is given by 
\begin{equation}
{\cal R}\sim 84\left(\frac{f_v}{0.1}\right)\left(\frac{\tau}{0.1}\right)
\left(\frac{T}{10^3\mathrm{K}}\right).
\end{equation}
In the isobaric mode, $\tau \sim 0.1$ from Fig. \ref{00}. 
Because the Raynolds number is much larger than unity, 
the dynamical condensation of the post-shock region 
is expected to be well described by 
the S-S solution with $\tau\ga 0.1$.

For the high-temperature case, 
the cooling rate of metal lines ($10^5$K $<T<$ $10^7$K) is given by
\begin{equation}
    \Lambda(T) \sim 2.2\times10^{29}\rho^2 T^{-0.6} \;\;\mathrm{ergs\;cm^{-3}\;s^{-1}}.
\end{equation}
We adopt $K=1.24\times 10^{-6} T^{5/2}$ ergs cm$^{-1}$ K$^{-1}$ s$^{-1}$ \citep{P53}. 
Using this formula and eqution (\ref{ray}), the Raynolds number is given by
\begin{equation}
{\cal R}\sim85\left(\frac{f_v}{0.1}\right)\left(\frac{\tau}{0.1}\right)
\left(\frac{T}{10^6\mathrm{K}}\right)^{0.1}.
\end{equation}
Because of ${\cal R}\gg1$, in the high-temperature case, 
the S-S solution with $\tau \ga 0.1$ is expected to 
well describe the dynamical condensation.

As the temperature decreases, 
${\cal R}$ decreases and becomes less than unity at a certain epoch.
Thereafter, our S-S solutions are invalid 
and the effect of dissipation becomes important.
\section{Summary}\label{summary}
We have provided S-S solutions describing the dynamical condensation 
of a radiative 
gas with $\Lambda - \Gamma\sim \rho^2 T^\alpha$. 
The results of our investigation are summarized as follows:

\begin{enumerate}
\item 
Given $\alpha$, a family of 
S-S solutions written as equation (\ref{den P time}) is found. 
The S-S solutions have the following two 
limits. The gas condenses almost isobarically for $\eta\sim 1$,
and the gas condenses almost isochorically for $\eta\sim 0$.
The S-S solutions exist for all values of $\eta$ 
between the two limits ($0<\eta<1$).
No S-S solutions are found in $\alpha>1$.
\item 
The derived S-S solutions are compared with the results of numerical 
simulations for converging flows. 
In the case with any scale of perturbation in the post-shock region,
our S-S solutions approximate 
the results of numerical simulations well in the high density limit.
In the post-shock region, smaller-scale perturbations grow faster and 
asymptotically approach to the S-S solutions for $\eta\sim 1$.
However, the difference of the growth rate between the isobaric mode
($\eta_\mathrm{eq}<\eta<1$) and 
the isochoric mode ($0<\eta<\eta_\mathrm{eq}$) is small.
Therefore, in actual astrophysical environments,
the S-S condensations for various $\eta$ are expected to occur simultaneously.
\end{enumerate}

In the S-S solutions presented in this paper, a plane-parallel geometry is
assumed. This situation is much simpler than 
the two-dimensional simulations that are currently being carried out 
involving TI, so the results are of limited use,
although the results may be relevant to code tests that are run on a
one-dimensional simulation.
The stability analysis on the 
present S-S solution may provide some insight into the multi-dimensional 
turbulent velocity field.
The effect of magnetic field and self-gravity may 
also be important in ISM but they are neglected in this paper.
These issues will be addressed in forthcoming papers.

\section*{Acknowledgments}
This work is in part supported by the 21st Century COE Program 
"Towards a New Basic Science; Depth and Synthesis" in Osaka University, 
funded by the Ministry of Education, Science, Sports and Culture of Japan.

%
\end{document}